\documentclass[aps,nofootinbib,superscriptaddress, showpacs,preprintnumbers,  nofootinbibt,onecolumn]{revtex4-2}
\usepackage{booktabs}
\usepackage[utf8]{inputenc}
\usepackage[T1]{fontenc}
\usepackage{amsmath}
\usepackage{amssymb}
\usepackage{graphicx}
\usepackage{caption}
\usepackage{subcaption}
\usepackage[most]{tcolorbox}
\usepackage{mathtools}
\usepackage{epsfig}
\usepackage{multirow}
\usepackage{eurosym}
\usepackage{dcolumn}
\usepackage{bm}
\usepackage{enumerate}
\usepackage{float}
\usepackage{epstopdf}
\usepackage{amsmath}
\usepackage{bm}
\usepackage{amsfonts}
\usepackage{amssymb}
\usepackage{graphicx}
\usepackage{alphalph,mathtools}
\usepackage{etoolbox}
\usepackage{color}
\usepackage{booktabs}
\usepackage{footnote}
\usepackage{makecell,tabularx}
\usepackage[leftcaption]{sidecap}
\usepackage{amssymb}
\usepackage{graphicx}
\usepackage{booktabs,dcolumn,caption}
\usepackage{rotating}
\usepackage{amsthm}
\usepackage{tikz}
\usepackage{caption}
\usepackage{subcaption}
\usepackage[most]{tcolorbox}

\usepackage{hyperref}
\hypersetup{colorlinks,citecolor=blue}
\hypersetup{colorlinks=true,linkcolor=red,filecolor=green,    urlcolor=blue}

\def\be{\begin{equation}}
	\def\ee{\end{equation}}
\def\bea{\begin{eqnarray}}
	\def\eea{\end{eqnarray}}

\begin{document}
	
	\title{An Eternal gravitational collapse in $f(R)$ theory of gravity and their astrophysical implications}
	\author{Annu Jaiswal}
	\email{annujais3012@gmail.com}
	
	\author{Rajesh Kumar}
	\email{rajesh.kumar@associates.iucaa.in}
     \email{rkmath09@gmail.com}
	\author{Sudhir Kumar Srivastava}
	\email{sudhirpr66@rediffmail.com}
	\affiliation{Department of Mathematics and Statistics, D.D.U.Gorakhpur University, Gorakhpur, India.}
	\author{Megandhren Govender}
	\email{megandhreng@dut.ac.za}
	\affiliation{Department of Mathematics, Faculty of Applied Sciences, Durban University of Technology, Durban 4000, South Africa}

	\begin{abstract}
In this work, we explore the eternal collapsing phenomenon of a stellar system (e.g., a star) within the framework of $f(R)$ gravity and investigate some new aspects of the continued homogeneous
gravitational collapse with perfect fluid distribution. The exact solutions of field equations have been obtained in an independent way by the parameterization of the expansion scalar ($\Theta$) governed by the interior spherically symmetric FLRW metric. We impose the Darmois junction condition required for the smooth matching of the interior region to the Schwarzschild exterior metric across the boundary hypersurface of the star. The junction conditions demand that the pressure is non-vanishing at the boundary and is proportional to the non-linear terms of $f(R)$ gravity, and the mass function $m(t, r)$ is equal to Schwarzschild mass $M$. The eight massive stars, namely $Westerhout 49-2, BAT99-98, R136a1, R136a2, WR 24, Pismis 24-1$, $\lambda- Cephei$,  and $\beta -Canis Majoris$ with their known astrophysical data (masses and radii) are used to estimate the numerical values of the model parameters which allows us to study the solutions numerically and graphically. Here we have discussed two $f(R)$ gravity models describing the collapse phenomenon.
The singularity analysis of models is discussed via the apparent horizon and we have shown that stars tend to collapse for an infinite co-moving time in order to attain the singularity (an eternal collapsing phenomenon).  We have also shown that our models satisfy the energy conditions and stability requirements for stellar systems.
		
	\end{abstract}
	
	\maketitle
	\tableofcontents

	\section{ Introduction}\label{sec1}
	The origins of modified gravity theories may be traced back to the necessity to explain some observable events that were not adequately accounted for by classical General Relativity (GR theory). Although GR theory has been enormously successful in describing a wide range of gravitational phenomena, there are still unsolved concerns and unexplained results that have prompted the development of alternative gravity theories.  The main driving factors behind modified gravity theories are some inconsistencies at ultraviolet scales (e.g. the initial singularity, the quantum gravity issue) and infrared scales (e.g. accelerating expansion of the universe, concordance problem, flatness problem, galaxy rotation curves, large-scale structure, massive star formation) that strongly suggest that Einstein’s approach should be revised or at least extended \cite{Capozziello2012}. Observations of distant supernovae in the late 1990s revealed that the universe's expansion is speeding, contrary to what was predicted based on gravitational interaction between matter. This prompted the proposal of dark energy, but it also generated curiosity about changing some basics of GR theory itself. In $f(R)$ gravity, Einstein's GR theory is modified by introducing a function of the Ricci scalar, $R$, into the Einstein-Hilbert action which incorporates the dynamics of a physical system\cite{Buchdahl1970}-\cite{Barrow1983}. This modification leads to modified field equations that govern the behavior of gravity. The origins of $f(R)$ gravity can be traced back to the 1970s when researchers began exploring alternative theories of gravity. The specific form of $f(R)$ gravity was proposed by Brans and Dicke \cite{Brans1961} as a scalar-tensor theory of gravity. In the 21st century, $f(R)$ gravity received renewed interest due to its potential to explain the observed acceleration of the universe's expansion without the need for dark energy. In particular, the theory gained attention after the discovery of cosmic acceleration by the Supernova Cosmology Project and the High-Z Supernova Search Team in 1998 \cite{Riess1998}. For a thorough examination of the motivation, the validity of various functional forms $f(R)$, applications, as well as shortcomings of these gravity theories, have been extensively analyzed \cite{Sotiriou2010}-\cite{R11}.
\par
Theoretical astrophysicists face a significant challenge in understanding the end state of a massive collapsing star. The question of the end state of collapsing stars with ultimate masses of 5$M_\odot$ or greater, after exhausting their nuclear fuels, has been a critical astronomical and astrophysics problem for decades\cite{book}. The nature of the initial data from which the collapse results is intrinsically dependent upon the possible outcome of continuous gravitational collapse leading to either a naked singularity or a black hole \cite{book1}-\cite{book3}. A continuous collapse without any equilibrium state has also been studied in some recent investigations; this phenomenon is referred to as "eternal gravitational collapse." \cite{ Mitra2010}-\cite{RJ23a}. To understand the potential end state of a continued gravitational collapse, it is essential to examine the dynamics of stellar collapse scenarios within the framework of gravitational theories. In the work\cite{R12}, the collapse of a star within modified gravity was addressed, and it was demonstrated that a class of $f(R)$ theories can result in curvature singularity in such a scenario. A general overview of self-gravitating dust collapse processes under $f(R)$ gravity was discussed in \cite{R13}. An extensive numerical analysis of the formation of black holes in spherical scalar collapse in these theories was carried out \cite{R14}. The phenomenon of gravitational collapse in the light of $f(R)$ gravity has gained much attention in the past few decades (\cite{R13}-\cite{R14c} and references therein).
\par
Motivated by our recent work\cite{EPJC}, the present study concerns the investigation of the continued collapsing phenomenon (ECO) in f(R) gravity theory. We considered a simple FLRW metric to start with, so in that sense, it is not very general, but we obtained an exact solution for a collapsing model that is valid for a reasonable domain of the theory. We discussed the possibility of the formation of a continuing collapsing phenomenon, i.e., no real space-time singularity, as the end product of the collapse is found to be dependent on the initial conditions. Here, we have discussed two models of $f(R)$ gravity, namely: (a) $f(R) = R + \alpha R^2$  and (b) $R = R^{1+ \beta}$  which play a significant role in the study of the gravitational collapse. For the astrophysical implications of our models, we considered the data of known massive stars and then discussed the possible dynamics and final stage of the collapsing system. 
\par
The paper is organized as follows: after the Introduction, we present the governing Einstein field equations (EFEs) for FLRW space-time metric with perfect fluid distributions in section II. In section III, we consider the exterior region of the system described by the Schwarzschild metric and then obtain the matching conditions for $f(R)$ gravity. In section IV we discussed the physical analysis of a collapsing stellar system by using a parameterization of the expansion scalar and to obtain exact solutions of EFEs. In section V, we examined the Physical viability of the obtained solutions with two $f(R)$ models for known stellar data. Also, we discussed the energy conditions and stability analysis of our model. In Section VI, we discussed the phenomenon of an eternal collapsing system which reveals a new aspect of the end-state of stars.  We conclude with a discussion of our findings in section VII.
	.
	
	
\section{Metric formalism and Basic equations in $f(R)$ gravity}\label{sec2}
The present study examines the spherically symmetric gravitational collapse of a stellar system using the FLRW space-time as its interior geometry 
	\begin{equation}	
		ds_{-}^2 = -dt^2 + a^2(t)dr^2 + \mathcal{R}^2 d \Omega^2
		\label{eq6}
	\end{equation}	
	where the coordinate is $x^i_{-} = (t,r,\theta,\phi)$, $d\Omega^2 \equiv d\theta^2 + \sin^2\theta  d\phi^2$ is the metric on two-sphere and $\mathcal{R} = r a(t)$ is the areal radius.
 \par

The f(R) theory of gravity is a notable and remarkable theory that generalizes the Einstein-Hilbert action using the Ricci scalar function\cite{ Buchdahl1970}. Researchers are exploring the variation in the nature of a collapsing star in terms of the $f(R)$ gravity and can be seen that it not only prevents the central singularity in the collapsing process but also slows the process \cite{ R12}. The modification in $f(R)$ gravity is derived by generalizing the Lagrangian in the Einstein-Hilbert action so that the Ricci scalar $R$ is replaced by a function $f(R)$ and on the basis of this, the action in $f(R)$ gravitational field theory is expressed as\cite{R14a}
	\begin{equation}	
		S = \int  \left[ \frac{f(R)}{2\mathcal{X}} + \mathcal{L}_m \right] \sqrt{-g} d^4x
		\label{eq1}
	\end{equation}	
	where $f(R)$ is any arbitrary function of the scalar curvature $R= g^{ij} R_{ij}$, where the Ricci tensor is given by $R_{ij} = R^{h}_{i j h}$, $g$ is the determinant of the metric tensor $g_{ij}$ ($i, j = 0, 1, 2, 3$), $\mathcal{X}$ is the  gravitational coupling constant and $\mathcal{L}_m$ is the Lagrangian of matter fields.
 \par
 Varying the action (\ref{eq1}) with respect to the metric tensor $g_{ij}$ yields the following field equation
	\begin{equation}
		F(R) R_{ij} - \frac{1}{2} f(R) g_{ij} - [\triangledown_{i} \triangledown_{j}-  g_{ij} \square] F(R) = \mathcal{X}  T_{ij}
		\label{eq2}
	\end{equation}
	where $F(R) = \frac{df}{dR}$, $\square \equiv \triangledown^{i} \triangledown_{i} $ with $\triangledown_{i}$ represents the covariant derivative.
	Writing the above equation in the form of Einstein tensor
	\begin{equation}
		G_{ij} = \frac{\mathcal{X} }{F(R)}\left(T^m_{ij}  +  T^{D}_{ij}\right)
		\label{eq3}
	\end{equation}
	where
	\begin{equation}
		T^{D}_{ij} = \frac{1}{\mathcal{X}} \left[ F(R) R_{ij} - \frac{1}{2} f(R) g_{ij} - \left( \triangledown_{i} \triangledown_{j}-  g_{ij} \square \right) F(R) \right]
		\label{eq4}
	\end{equation}
	
	and $T^m_{ij}$ is the energy-momentum tensor describing fluid distribution inside the stellar system which is  considered to be a perfect fluid 
	\begin{equation}
		T^m_{ij} = (p+\rho) V_i V_j + p g_{ij}
		\label{eq5}
	\end{equation}
	where $\rho$ and $p$ are respectively the energy density and pressure of the fluid distribution. 
	
	Thus, the $f(R)$ gravity field equations(\ref{eq3}) for metric(\ref{eq6}) yields the following equations
	
	\begin{equation}
		 \mathcal{X}  \rho (t)= 3 F \frac{\ddot{a}}{a} - 3  F^{'} \dot{R}\frac{\dot{a}}{a} - \frac{f}{2}
		\label{eq7}
	\end{equation}
	
	\begin{equation}
	\mathcal{X}p = -F\left(\frac{2\dot{a}^2}{a^2} + \frac{\ddot{a}}{a}\right) + 2  F^{'} \dot{R} \frac{\dot{a}}{a} +  F^{'} \ddot{R}+  F^{''}  \dot{R}^2   + \frac{f}{2}
		\label{eq8}
	\end{equation}
	where dash $(')$ and dot $(.)$ represent differentiation with respect to Ricci scalar $R$ and time $t$ respectively.
 \par
	Since for the collapsing configuration $ \frac {\dot{\mathcal{R}}}{\mathcal{R}} < 0$ and the collapsing rate of the system is described by the expansion-scalar($\Theta$) 
	\begin{equation}
		\Theta = V^i_{;i} = 3\frac{\dot{\mathcal{R}}}{\mathcal{R}}
		\label{eq9}
	\end{equation}
	The mass function $m(t,r)$ for the spherically symmetric collapsing system at any instant $(t,r)$ is given by \cite{me70}   \\
	\begin{equation}
		m(t,r) = \frac{1}{2} R \left(1 + R_{,i} R_{,j} g^{ij}\right) = \frac{1}{18} \Theta^2 R^3 
		\label{eq10}
	\end{equation}	
	where the comma (,) denotes the partial derivative.
 \par
\textbf{The Kretschmann Curvature}
\par
The smoothness of a space-time is an important aspect to consider when studying its geometry. Smooth space-time simply means that it has regular curvature invariants that are finite at all spacetime sites, or it has curvature singularities, at least one of which is infinite. The Kretschmann scalar curvature (KS) widely used technique to evaluate the smoothness of space-time in various situations (which is also known as the Riemann tensor squared) and is defined by\cite{SO02} \cite{1992}
	\begin{equation}
		\mathcal{K} = R_{i j k \delta} R^{i j k \delta}
		\label{eq13}
	\end{equation}
	where $R_{i j k \delta}$ denotes Riemann curvature tensor. For the metric (\ref{eq6}), we have
	\begin{equation}
		\mathcal{K} = \dfrac{12 \left(\dot{a}^4 + a^2 \ddot{a}^2\right)}{a^4}
		\label{eq14}
	\end{equation}  
	The space-time singularity is identified when the  Kretschmann curvature $\mathcal{K}$ diverges uniformly.
	
\section{Matching conditions}\label{sec3}
According to Jebsen-Birkhoff's theorem, the exact solution of vacuum Einstein field equations describing the gravitational field exterior to a spherically symmetric system is the Schwarzschild metric. Here we consider the exterior region of the star to be described by the Schwarzschild geometry and the metric of it is given by
	\begin{equation}
		ds^{2}_{+} = -\left(1-\frac{2M}{\mathbf{r}}\right) dT^{2} + \frac{d\mathbf{r}^{2}}{\left(1-\frac{2M}{\mathbf{r}}\right)} + \mathbf{r}^{2} d\Omega^{2}
		\label{eq11}
	\end{equation}
	where $M$ represents the Newtonian mass of the star (also known as Schwarzschild mass) and the coordinate of exterior space-time is $ x^{i}_{+}=(T, \mathbf{r}, \theta, \phi)$.
	\par
	In this section, we derive the matching conditions(junction conditions) for $f(R)$ theories of gravity: the generalized Darmois-Israel junction conditions. These junction conditions are necessary to study the stellar models where a vacuum region surrounds a stellar object in equilibrium. Given its importance for our understanding of gravitational collapse with proper matching conditions in non-linear $f(R)$ theories of gravity, it can be used as boundary conditions for any stellar system. 
 \par
 Let us consider the junction conditions for the smooth matching of the interior manifold $ds^2_{-}$ (metric(\ref{eq6}) as considered above) with the exterior manifold $ds^2_{+}$   across hypersurface $\Sigma$.  The boundary hypersurface $\Sigma$ divides the spherically symmetric stellar system into the interior $(ds^{2}_{-})$ and the exterior $(ds^{2}_{+})$ space-time region.
	\par
	Consider the instrinsic metric $ds^2$ over the boundary  hypersurface $\Sigma$ as 
	
	\begin{equation}
		ds^2_\Sigma = -d\tau^2 + \mathcal{Y}(\tau)^2 \left( d\theta^2 +\sin\theta^2 d\phi^2  \right)
		\label{eq11a}
	\end{equation}
where $\xi^i = (\tau, \theta, \phi)$ are coordinate on boundary $\Sigma$.
\par
The junction conditions in $f(R)$ theory of gravity are different from those of general relativity\cite{jmm13}-\cite{deruelle et al08}. Besides the known Darmois-Israel junction conditions in GR (continuity of metric and extrinsic curvature), the non-linear metric $f(R)$ gravity requires two additional constraints to be satisfied so as to make a given matching possible.
\par
	\begin{itemize}
	\item First junction condition: the continuity of metric at $\Sigma$ i.e. 
	\end{itemize}
\begin{equation}
	(ds^2_{-})_\Sigma = (ds^2)_\Sigma = (ds^2_{+})_\Sigma 
	\label{eq11b}
\end{equation}
which gives
\begin{equation}
dt \overset{\Sigma}{=} \left(1-\frac{2M}{R}\right) dT \overset{\Sigma}{=} d\tau, \quad~ ra(t) \overset{\Sigma}{=} \mathbf{r} \overset{\Sigma}{=}  \mathcal{Y}(\tau)
\label{jun1}
\end{equation}

\begin{itemize}
	\item Second junction condition: the continuity of the extrinsic curvature at $\Sigma$, i.e.
\end{itemize}
\begin{equation}
K^{+}_{ij} \overset{\Sigma}{=}  K^{-}_{ij} 
\label{eq11c}
\end{equation}
	
	where $K_{ij}$ denotes the extrinsic curvature defined by
	\begin{equation}
		K^{\pm}_{ij} = -N^{\pm}_{k} \left(\frac{\partial^{2} x^{k}_{\pm}}{\partial \xi^{i} \partial \xi^{j}} + \Gamma^{k}_{\alpha \beta}\frac{\partial x^{\alpha}_{\pm}}{\partial \xi^{i}} \frac{\partial x^{\beta}_{\pm}}{\partial \xi^{j}}\right)
		\label{eq11d}
	\end{equation}
	where $k, \alpha, \beta = 0,1,2,3$ and $N^{\pm}_{k}$ are components of  unit normals to $\Sigma$ in the coordinates $x^i_{\pm}$ and are given by
	\begin{equation}
		N^{+}_{k} \overset{\Sigma}{=} (-\frac{d\mathbf{r}}{d\tau},\frac{dT}{d\tau}, 0, 0)
		\label{eq11e}
	\end{equation}
	\begin{equation}
		N^{-}_{k} \overset{\Sigma}{=} (0,a(t),0,0)
		\label{eq11f}
	\end{equation}
	
The components of the extrinsic curvature $K_{ij}$ are
\begin{equation}
K^{-}_{\tau\tau} \overset{\Sigma}{=} 0
\label{jun2}
\end{equation}
\begin{equation}
K^{-}_{\theta\theta} \overset{\Sigma}{=} r a(t) = K^{-}_{\phi\phi} \csc^2\theta
\label{jun3}
\end{equation}

\begin{equation}
K^{+}_{\tau\tau} \overset{\Sigma}{=} \left[ \frac{d\mathbf{r}}{d\tau}\frac{d^2 T}{d\tau^2}-\frac{dT}{d\tau}\frac{d^2 \mathbf{r}}{d\tau^2}+\frac{3M}{\mathbf{r}(\mathbf{r}-2M)}\frac{dT}{d\tau} (\frac{d\mathbf{r}}{d\tau})^2-\frac{M(\mathbf{r}-2M)}{\mathbf{r}^3} (\frac{dT}{d\tau})^3 \right]
\label{jun4}
\end{equation}
\begin{equation}
K^{+}_{\theta\theta} \overset{\Sigma}{=} (\mathbf{r}-2M)\frac{dT}{d\tau} = K^{+}_{\phi\phi} \csc^2\theta
\label{jun5}
\end{equation}
Substituting Eqs.(\ref{jun3}) and (\ref{jun5}) into condition $K^{-}_{\theta\theta} \overset{\Sigma}{=} K^{+}_{\theta\theta} $  and using (\ref{jun1}) and (\ref{eq10}), we obtain
\begin{equation}
	m(t,r)  \overset{\Sigma}{=} M
	\label{junction1}
\end{equation} 
This is just like in general relativity which shows that collapsing mass $m(t, r)$ must be equal to the Schwarzschild mass $M$ on boundary $\Sigma$.
\par

Again the conditions $K^{-}_{\tau\tau}  \overset{\Sigma}{=} K^{+}_{\tau\tau}$ gives
\begin{equation}
\frac{d\mathbf{r}}{d\tau}\frac{d^2 T}{d\tau^2}-\frac{dT}{d\tau}\frac{d^2 \mathbf{r}}{d\tau^2}+\frac{3M}{\mathbf{r}(\mathbf{r}-2M)}\frac{dT}{d\tau} (\frac{d\mathbf{r}}{d\tau})^2-\frac{M(\mathbf{r}-2M)}{\mathbf{r}^3} (\frac{dT}{d\tau})^3 =0
\label{jun6}
\end{equation}	
Now, substituting the values from Eq.(\ref{jun1}) and using FEE (\ref{eq8}) into (\ref{jun6}), we obtain
\begin{equation}
\mathcal{X} p + \frac{1}{2} (f-R F) + 2\frac{\dot{a}}{a} F^{'} \dot{R} + F^{'} \ddot{R} +\dot{R}^2 F^{''} \overset{\Sigma}{=} 0
\label{junction2}
\end{equation}
From Eq.(\ref{junction2}) we see that the pressure is non-vanishing at the boundary $\Sigma$ but instead is proportional to the non-linear terms $f(R)$ gravity\footnote{For $f(R) =R$ and then condition(\ref{junction2}) reduces to $p=0$, which is well-known junction condition in classical general GR theory}. These extra non-linear terms may appear due to the higher order curvature geometry of the collapsing sphere in $f(R)$ gravity.

\begin{itemize}
		\item Third junction condition: the continuity of the Ricci scalar at $\Sigma$, i.e.
		\begin{equation}
			R^{-} \overset{\Sigma}{=} R^{+}
			\label{eq11g}
		\end{equation}
	\end{itemize}
	
	\begin{itemize}
		\item Fourth junction condition: the continuity of the normal derivative of the Ricci scalar at $\Sigma$, i.e.
		
		\begin{equation}
			N^k_{-} \frac{\partial R^{-}}{\partial x^{k}_{-}} \overset{\Sigma}{=} N^k_{+} \frac{\partial R^{+}}{\partial x^{k}_{+}}
			\label{eq11h}
		\end{equation}
	\end{itemize}
	The existence of third and fourth junction conditions in metric $f(R)$ gravity is due to the fact that these theories propagate an additional scalar degree of freedom in comparison with GR, and that this scalar mode is intimately related to the Ricci scalar \cite{sotirio10}-\cite{casado et al22}. Since Ricci scalar ($R^{+}$) vanishes for the exterior Schwarzschild metric (\ref{eq11}), then the conditions (\ref{eq11g})-(\ref{eq11h})  are to be satisfied for the metric of the internal manifold at the hypersurface $\Sigma$ identically.
 \par
	Theoretically, the boundary conditions are used to estimate the numerical value of arbitrary constants. Therefore, in our studies, we can determine the arbitrary constant and model parameters by using the aforementioned conditions (\ref{junction1}) and (\ref{junction2}).


 \section{Physical analysis of the collapsing stellar system}\label{sec4}
\subsection{Parametrization of Expansion scalar ($\Theta $)}\label{sec4a} 
Here we discuss a constraint equation for the complete analysis of EFEs (\ref{eq7})-(\ref{eq8}) by considering the parametrization of expansion scalar \cite{EPJC}
\begin{equation}
	\Theta = -\frac{16 \gamma  t^3}{15 \left(\frac{8 \gamma  t}{5}+1\right)},~\quad~\gamma > 0
	\label{eq15}
\end{equation}
where $\gamma$ is a model parameter to be determined by using the observational data of some known massive stars(see Table-\ref{table1}).
\par

\begin{figure}[h]
		\includegraphics[scale=0.80]{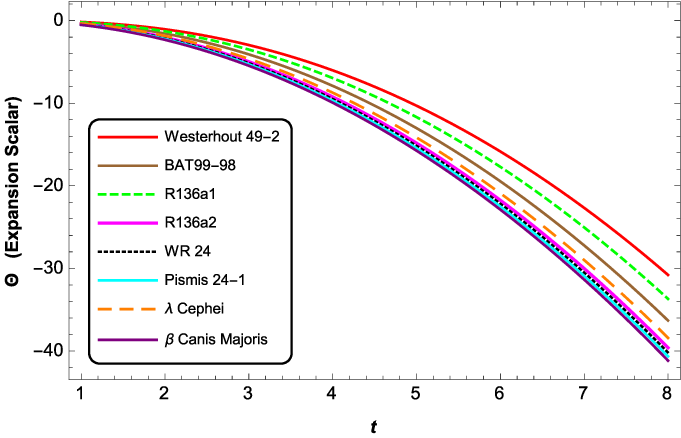}
		\caption{Collapsing configuration: show the plot of the "variation of expansion scalar $\Theta$" v/s time coordinate $t$ for the massive stars, with model parameter(s) $\gamma$ given in Table-\ref{table1}. The unit of time $t$ is Myr (Million years).}
		\label{fig1}
	\end{figure}

The physics of choosing parametrization(\ref{eq15}) are as follows- in the evolution of the stellar system (e.g., a star) due to nuclear fusion in the core, it loses its equilibrium stage and starts to collapse under its own gravity. During the collapse process, the internal thermal pressure (which is due to nuclear fusion of hydrogen or helium ) decreases, and then the external gravitational pressure (which is due to the mass of the star) dominates over it. At the same time, the collapse rate of the star increases which tends to draw matter inward towards the center of the stellar configuration. According to GR, in the collapsing system, two kinds of motion (velocities) occur namely $\frac{\dot{\mathcal{R}}}{\mathcal{R}}$ which measures the variation of areal radius $\mathcal{R}$ per unit proper time and, another $(\delta l)^{.}$, the variation of the infinitesimal proper radial distance, $(\delta l)$  between two neighboring fluid particles per unit proper time \cite{EPJC}\cite{herrera2009}. The expansion scalar $\Theta$ is defined as the rate of change of elementary fluid particles which describes the collapsing rate of the fluid distribution in a stellar system.  Since the FLRW homogeneous gravitational collapse requires the motion of fluid particles to be uniform, independent of $r$ and fluid motion speed up towards the center. Consequently, the expansion scalar ($\Theta$) increases with $t$  in a collapsing scenario (see figure \ref{fig1}).


 \subsection{Exact solution of field equations}\label{sec4b}
	We consider the parametrization (\ref{eq15}) as an additional constraint to obtain the solution of field equations (\ref{eq7})-(\ref{eq8}). Using of eqs.(\ref{eq15}) into (\ref{eq9}) and integrating, we obtain
	\begin{equation}
		a(t) = k e^ {-\frac{16}{9} \gamma  \left(\frac{25 t}{512 \gamma ^3}-\frac{5 t^2}{128 \gamma ^2}+\frac{t^3}{24 \gamma } -\frac{125 \log (8 \gamma  t+5)}{4096 \gamma ^4}\right)}
		\label{eq17}
	\end{equation}
	where $k$ is an integrating constant.  In order to determine the value of  $k$, we use the boundary condition (\ref{junction1}).
	\par
	Assuming that the star begins to collapse initially at $t=t_{0}$ from boundary $\Sigma$ at $r=r_0$. Then by Using the eqs. (\ref{eq10}) and (\ref{eq17})  into eq.(\ref{junction1}) yield
	\begin{equation}
		k =  \frac{3 {\left(\frac{3}{2}\right)^{\frac{1}{3}} M^{\frac{1}{3}} e^{\frac{{t_0} \left(64\gamma^2 {t_0}^2-60 \gamma{t_0}+75\right)}{864 \gamma^2}} (8 \alpha {t_0}+5)^{\frac{1}{3} \left(2-\frac{125}{768 \gamma^3}\right)}}}{4 \gamma^{\frac{2}{3}}{r_0} {t_0}^2}
		\label{eq17a}
	\end{equation} 
	where $(t_0, r_0)$ is the initial coordinate of the star when it starts to collapse. Thus, we have from (\ref{eq17}) that \cite{EPJC}
	\begin{equation}
		a(t)  = \frac{ {(\frac{81 M}{2})}^{\frac{1}{3}} (8\gamma  {t_0}+5)^{\frac{2}{3}-\frac{125}{2304\gamma^3}}}{4 \gamma^{\frac{2}{3}} {r_0} {t_0}^2}  e^{\frac{375 \log (8 \gamma  t+5)-8 \gamma  \left(64 \gamma ^2 t^3-60 \gamma  t^2+75 t+{t_0} \left(-64 \gamma ^2 {t_0}^{2} + 60 \gamma  {t_0} -75 \right)\right)}{6912 \gamma ^3}}
		\label{eq19}
	\end{equation}

	Thus, using eq.(\ref{eq19}) into (\ref{eq7})-(\ref{eq8}),  we obtain the physical parameters 
	\begin{equation}
		\mathcal{X} \rho =\frac{16 \gamma  t^2}{27  (5 + 8 \gamma  t)^2} \left[9 t (8 \gamma  t+5)  F^{'} \dot{R} +  \left(16 \gamma  t^4-144 \gamma  t-135\right)F \right]  -\frac{f}{2}
		\label{eq20}
	\end{equation}

		\begin{equation}
			\mathcal{X} p = \frac{16 \gamma  t^2}{27  (5 + 8 \gamma  t)^2} \left(45 + 48 \gamma t - 16 \gamma t^{4}\right) F - \frac{1}{9(5 + 8 \gamma  t)} \Bigl[32 \gamma t^3 F^{'} \dot{R} - 9 (5 + 8 \gamma  t) (F^{''} \dot{R}^2 + F' \ddot{R})\Bigr] + \frac{f}{2} 
			\label{eq21}	
		\end{equation}
	Using eq. (\ref{eq19}) into eq.(\ref{eq10}) gives\cite{EPJC}
	\begin{equation}
		m(t,r) =\frac{M}{r_{0}^{3}  t_{0}^{6}} (8 \gamma  t + 5)^{2-\frac{125}{768 \gamma ^3}} r^3 t^{6} (8 \gamma  t + 5)^{\frac{125}{768 \gamma ^3}-2} \\ e^{\frac{-64 \gamma ^2 t^{3} + 60 \gamma  t^{2} - 75 t + t_{0} \left(64 \gamma ^2 t_{0}^{2}-60 \gamma  t_{0} + 75\right)}{288 \gamma ^2}}
		\label{eq22}
	\end{equation}
	
	Also from eq.(\ref{eq14}), the Kretschmann curvature
	\begin{equation}
		\mathcal{K} = \frac{1024 \gamma^2 t^4}{2187 (8 \gamma t+5)^4} ( 512 \gamma^2 t^8 -  4608 \gamma^2 t^5- 4320\gamma  t^4  +20736 \gamma^2 t^2	+ 38880 \gamma t+18225 )
		\label{eq25}
	\end{equation}
	The collapsing acceleration($\frac{\ddot{a}}{a}$) is obtained from eq. (\ref{eq19}) that\cite{EPJC}
	\begin{equation}
		\frac{\ddot{a}}{a}=\frac{16  \gamma  t^2 \left(16  \gamma  t^4-144 \gamma  t-135\right)}{81 (8  \gamma t+5)^2}
		\label{eq26}
	\end{equation}
	\begin{figure}[h]
	\centering
	\includegraphics[scale=0.75]{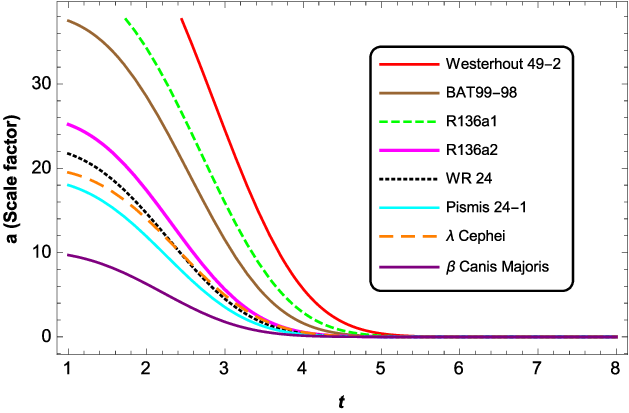}
	\caption{Show the plot of the "variation of scalar factor $a$" v/s  time coordinate $t$ for the massive stars }
	\label{fig2}
\end{figure}
	\begin{figure}[h]
		\centering
		\includegraphics[scale=0.70]{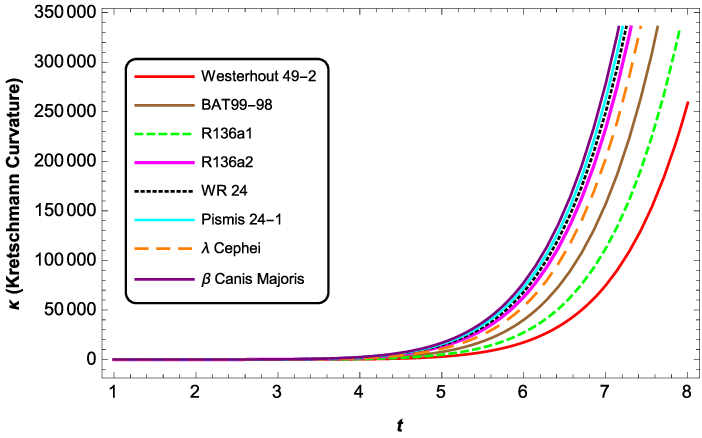}
		\caption{Show the plot of "Kretschmann scalar $\mathcal{K}$" v/s  time coordinate $t$ for the massive stars}
		\label{fig3}
	\end{figure}
 \begin{figure}[hbt!]
	\centering
	\includegraphics[scale=0.75]{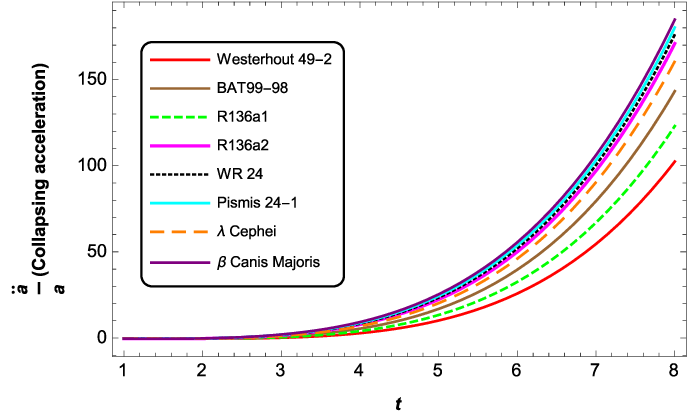}
	\caption{ Show the plot of "collapsing acceleration $\frac{\ddot{a}}{a}$" v/s  time coordinate $t$ for the massive stars, with model parameter(s) $\gamma$ given in Table \ref{table1}.}
	\label{fig4}
\end{figure}
\\
	It can be seen from eqs.(\ref{eq19})-(\ref{eq26}) that all the quantities namely $a$, $\,p$, $\rho $, $m$, $\mathcal{K}$ and $\frac{\ddot{a}}{a}$ are obtained in terms of stellar mass $M$ and $\gamma$. Therefore, the solution is applicable to the studies of the known stars whose masses and radii are given \cite{ob1}-\cite{ob7b}. By knowing model parameters $\gamma $, one can discuss the physical analysis of the models and explore their astrophysical significance.
	\par
	Using the observational data points: mass $(M)$ and radius $(\mathcal{R}_0)$ of known massive stars we have estimated the numerical value of model parameter $\gamma$ which are summarised in Table-\ref{table1} \cite{EPJC}, where $\mathcal{R}_0 = \mathcal{R}(t_0,r_0)$, assuming the initial coordinate $(t_0, r_0) = (1, 1)$, Solar mass $M_\odot$ and solar radius $ R_\odot$ to be unit.
	\begin{table}[hbt!]
		\caption{Numerical values of model parameter $\gamma$ for the known  masses $M$ and radii $R_0$ of eight known massive stars.}
		\setlength{\tabcolsep}{3
			\tabcolsep}
		\centering
		\begin{tabular}{*{4}{c}}
		\toprule
			\textbf{Massive Stars} & \textbf{$M (M_\odot)$} & \textbf{$ R_0 (R_\odot)$} & \textbf{$\gamma$}\\
			\midrule
			Westerhout 49-2 \cite{ob1} &	250	& 55.29 &	0.202544 \\ 
			BAT99-98 \cite{ob2}&	226 &	37.5 &	0.446334\\
			R136a1 \cite{ob3}	&196&	42.7&	0.293187\\
			R136a2\cite{ob4}&	151	&25.2&	1.0119\\
			WR 24 \cite{ob5}\cite{ob6}&	114&	21.73&	1.27352\\
			Pismis 24-1 \cite{ob6a}\cite{ob6b} & 74 & 18 & 1.58239\\
			$\lambda$ Cephei\cite{ob7} &	51.4&	19.5&	0.704378\\
			$\beta$ Canis Majoris \cite{ob7a}\cite{ob7b} &	13.5 & 9.7 & 2.1404 \\
			\bottomrule
		\end{tabular}
		\label{table1}
	\end{table}
\section{Physical viability of the proposed solutions via $f(R)$ models}\label{sec8}

This section delved into the physical behavior of the collapsing stellar system, highlighting its importance for its credibility and stability. We study the physical affirmation of the proposed solutions for the collapsing massive stars via energy density, pressure, energy conditions, and stability analysis under two particular $f(R)$ models that are feasible and well-behaved in gravitational theory. Furthermore, the graphical illustrations are also discussed in detail.
Here, we consider two  forms of $f(R)$ models which are given by
\par
\textbf{Model-1:} $f(R) = R + \alpha R^2$ \cite{Starobinsky1980},~ \quad\quad~ \textbf{Model-2:} $f(R) = R^{1+\beta}$ \cite{bh08}
\par
where $\alpha$ and $\beta$ are model parameters to be determined by junction condition (\ref{junction2}) for known stellar data (see table-\ref{table2}).\\
The first one is known as Starobinsky’s model in the quadratic form of the Ricci scalar $R$. He proposed a model to identify the growth of early and late-time accelerating universes.  Most works introduce scalar curvature formation as a potential candidate for dark energy \cite{sv22}- \cite{oo19}. The second model was proposed for describing test particle motion around galaxies and uses minimal deviations from classical general relativity and modified gravity to explain galactic dynamics without dark matter \cite{bh08}-\cite{sv22}. There are qualitative results presented in the literature in $f(R)$ gravity models that are consistent in describing the realistic configuration of stellar systems in astrophysical backgrounds \cite{R12}-\cite{R14c}, \cite{gn14}-\cite{Chakrabarti2016}.
\par
For model-1, using $f(R)= R + \alpha R^2$ into eqs.(\ref{eq20})-(\ref{eq21}), the energy density and pressure become
\begin{align}
\mathcal{X} \rho &=	\frac{256 \gamma ^2 t^4 \left[t^2 \left(16 \gamma  t \left(64 \alpha  \gamma  t^2+60 \alpha  t-4 \gamma  t-5\right)-25\right)-450 \alpha \right]}{27 (8 \gamma  t+5)^4}
		\label{eq27}
\end{align}
	
		\begin{multline}
		\mathcal{X} p = \frac{32 \gamma}{27 \left( 5 + 8 \gamma  t \right)^4} \Biggl[t^2 \Bigl[ 8 \gamma  t \Bigl\{-40 \gamma  t \left(24 \alpha   \left(t^3-15\right) t-2 t^3+21\right)\\
  +25 \left(t^3+306 \alpha   t-24\right)-64 \gamma ^2 t^2 \left(8 \alpha  \left(2 t^3-15\right) t-t^3+6\right)\Bigr\} -1125\Bigr] -13500 \alpha  \Biggr]
		\label{eq28}
	\end{multline}
For model-2, using $f(R)= R^{1+\beta}$ into eqs.(\ref{eq20})-(\ref{eq21}), we obtain
\begin{multline}
	\mathcal{X} \rho =	\frac{2^{4  + 5 \beta } 27^{-1-\beta} \gamma ^2 t^4}{(8 \gamma  t+5)^4}\Biggl(\frac{\gamma  t^2 \bigl( -135 +16   t (-9+2 t^3)\gamma \bigr)}{(8 \gamma  t+5)^2}\Biggr)^{-1+\beta }\Biggl[-6075 \beta  \left(-1+2 \beta \right) +     2160   t \Bigl( -9 \left(-1+\beta \right) \beta +  t^3 \left(1+ \beta + 4  \beta^2 \right) \Bigr) \gamma \\ + 128  t^2 \Bigl( 4 t^6 \left(-1 + \beta \right)-81 \left(-1+\beta \right) \beta + 18t^3\left(1+ \beta +4 \beta ^2 \right)\Bigr) \gamma ^2\Biggr]
	\label{eq29}
	\end{multline}
 
			\begin{multline}
		\mathcal{X} p =	\frac{2^{1+5 \beta }  27^{-1-\beta}}{t^2 (5+8 \gamma  t)^2 \left(135 + 16   t (9-2 t^3)\gamma \right)^2} \Biggl(\frac{\gamma  t^2 \left(-135 + 16 t \left(-9 +2 t^3 \right)\gamma\right)}{(5+8 \gamma  t)^2}\Biggr)^{\beta}  \\  	\Biggl[
		16 \gamma  t^4 \bigl(135+16 \gamma  t \left(9-2 t^3\right)\bigr)^2  \bigl(-45+ 8   t \left(t^3-6\right)\gamma\bigr) +54 \beta ^{3} \bigl(675-8 \gamma  t \left(  -135+ 60 t^3 + 8 t  \left(-9+8 t^3\right) \gamma\right)\bigr)^2 \\   -\beta  \Bigl(12301875+8 \gamma  t \Bigl(91125 (90+19 t^3) -97200   t \left(-135 -51 t^{3} + t^{6}\right)\gamma + 103680  t^{2} \left(81 + 54t^{3} + 2 t^{6}\right) \gamma ^2+ \\ 1024 t^3 \left( 2187 + 4 t^{3}\left(486 + 27 t^3 +4 t^9\right) \right)\gamma ^3 \Bigr)\Bigr) -3 \beta ^2 \left( -135 + 16   t (-9 + 2 t^3)\gamma\right)  \\
		\Bigl(30375+ 64 \gamma  t^4 \left( -2025 + 4 \gamma  t \left( -945 + +60 t^3 + 8 t (8 t^3-63) \gamma \right)\right) \Bigr)
		\Biggr]		
		\label{eq30}
	\end{multline}
 The nature of arbitrary parameters in $f(R)$ models is crucial as they should be suitable and compatible with any stellar model. These are the model parameters that determine the physical as well as the graphical characteristics of the model. Thus, in our study, we have estimated these model parameters numerically by using the known stellar data points (their masses and radii) with boundary conditions (\ref{junction2}). The numerical values of model parameters $\alpha$ and $\beta$ are summarized in Table-\ref{table2}.
	\begin{table}[hbt!]
		\caption {Numerical values of $f(R)$ gravity  models parameters $\alpha$ and $\beta$ for the eight known massive stars.}
		\setlength{\tabcolsep}{3
			\tabcolsep}
		\centering
		\begin{tabular}{p{5.2cm}p{2.6cm}p{2cm}}
			\hline\specialrule{0.3pt}{0pt}{0pt}
		\toprule
			\textbf{Massive Stars} & \textbf{$\alpha$}\newline (Model-1) &  \textbf{$\beta$}\newline (Model-2)\\
			\midrule
   \addlinespace
			Westerhout 49-2 \cite{ob1} &	0.669189 & 1.05293  \\ 
           \addlinespace
   
			BAT99-98 \cite{ob2}  &	0.06826  &	1.8533\\
   \addlinespace
			R136a1 \cite{ob3}	&	0.130593 &	1.33155\\
   \addlinespace
			R136a2\cite{ob4} &	0.0389343  &	4.27634\\ \addlinespace
   
			WR 24 \cite{ob5} \cite{ob6} &	0.0358584 &	5.60847\\ \addlinespace
   
			Pismis 24-1 \cite{ob6a}\cite{ob6b} &    0.0337622 &     7.30816  \\ \addlinespace
   
			$\lambda$ Cephei\cite{ob7} &	0.046761 &	2.86998\\ \addlinespace

			$\beta$ Canis Majoris \cite{ob7a}\cite{ob7b}  & 0.0317087 & 10.6298\\ \addlinespace
\bottomrule
		\end{tabular}
		\label{table2}
	\end{table}
\\

	\begin{figure}[hbt!]
		\centering
		\includegraphics[scale=0.76]{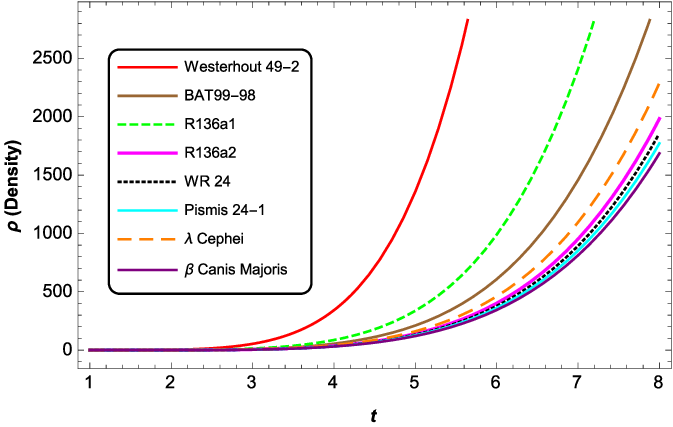}
		\caption{ Variation of "Energy density" v/s time $t$ for model-1 with values of  model parameters $\gamma$ and $\alpha$ given in table \ref{table1} and \ref{table2} respectively.}
		\label{fig4a}
	\end{figure} 
	\begin{figure}[hbt!]
		\centering
		\includegraphics[scale=0.79]{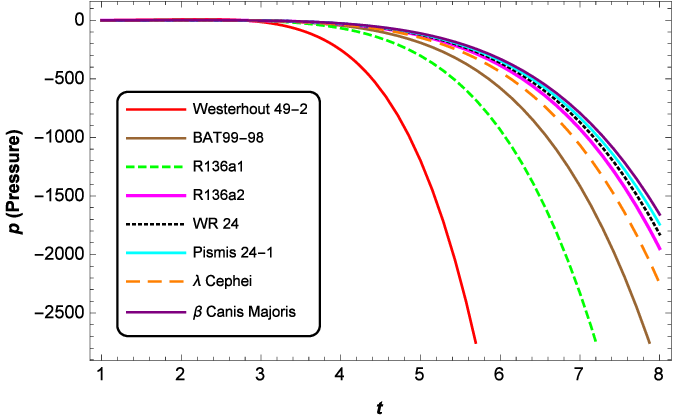}
		\caption{Variation of "Pressure" v/s time $t$ for model-1 with values of  model parameters $\gamma$ and $\alpha$ given in table \ref{table1} and \ref{table2} respectively.}
		\label{fig5}
	\end{figure}
 \begin{figure}[hbt!]
		\includegraphics[scale=0.75]{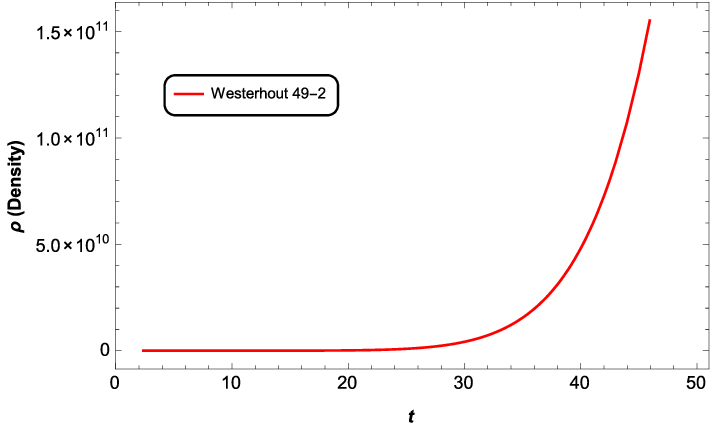}
		\caption{Variation of "Energy density" v/s time $t$ for model-2 with values of model parameters $\gamma$ and $\beta$ given in table \ref{table1} and \ref{table2} respectively.}
		\label{fig6}
	\end{figure}
	\begin{figure}[hbt!]
		\includegraphics[scale=0.80]{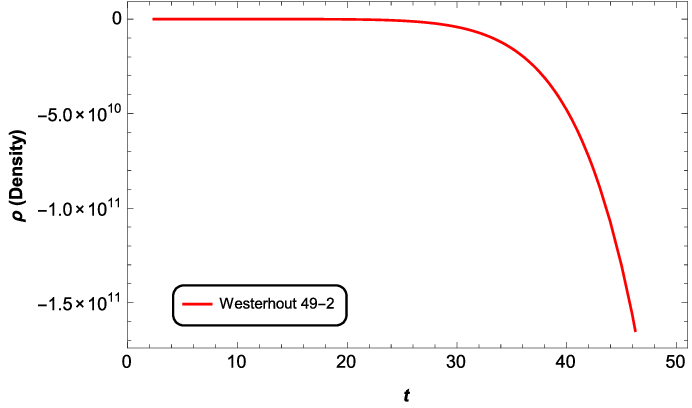}
		\caption{ Variation of "Pressure" v/s time $t$ for model-2 with values of model parameters $\gamma$ and $\beta$ given in table \ref{table1} and \ref{table2} respectively.}
		\label{fig7}
	\end{figure} 

 Figures-\ref{fig4a} and \ref{fig6} depict the dynamical evolution in energy density during the collapsing configuration\footnote{The graphs illustrating the variation of energy density and pressure for eight massive stars in model-2 are overlapping, therefore single diagram for "Wasterhout" is displayed here in order to provide a clear graphical representation.}.  In both respective plots the $f(R)$ models show a steady increase in energy density with time, its nature being positive and non-singular at all interior points of the stellar distribution.  It is noted here that, especially in the model-2 plot, the energy density is higher near the center as compared to the first model that actually predicts the ultra-dense stellar model. In figures \ref{fig5} and \ref{fig7}, we examined the profile of pressure in collapsing stars for both $f(R)$ models. It can be eminently seen from both graphs that the pressure evolution is increasing in nature and each plot of the pressure for both $f(R)$ models reveals negative evolution in the entire configuration of the collapsing star( where the negative sign indicates the direction of gravitational-pressure towards the center of the star).  It can also be observed that pressure is non-zero at the boundary surface of the star and has a maximal amount near the center($r \rightarrow 0$).
 \subsection{Energy Conditions}\label{energy}
This section discusses the physical characteristics of our collapsing stellar models, analyzing various energy conditions (EC). The study of astrophysical events, such as stellar collapsing objects, is significantly influenced by these conditions. The analysis has been conducted by ensuring the physical affirmations by satisfying various energy conditions. The conditions, including null (NEC), weak (WEC), dominant (DEC), and strong (SEC), must be validated and finite at each interior point of the entire stellar system \cite{e1}. The following constraints need to be fulfilled for EC:
\begin{align}
\mbox{\textbf{NEC}:} \quad \rho+p \geq 0
 \label{ec1}  
\end{align}
\begin{align}
 \mbox{\textbf{WEC:}} \quad~ \rho \geq 0,\quad~ \mbox{and} \quad ~\rho+p \geq 0
\label{ec2}   
\end{align}
\begin{align}
\mbox{\textbf{DEC:}} \quad~ \rho \geq \lvert p \lvert
\label{ec3}
\end{align}
\begin{align}
\mbox{\textbf{SEC:}} \quad \rho+p \geq 0,\quad~ \mbox{and} \quad~ \rho+ 3 p \geq 0
\label{ec4}
\end{align}

\begin{figure}[hbt!]
		\begin{subfigure}[a]{0.4\textwidth}
			{\includegraphics[width=8cm]{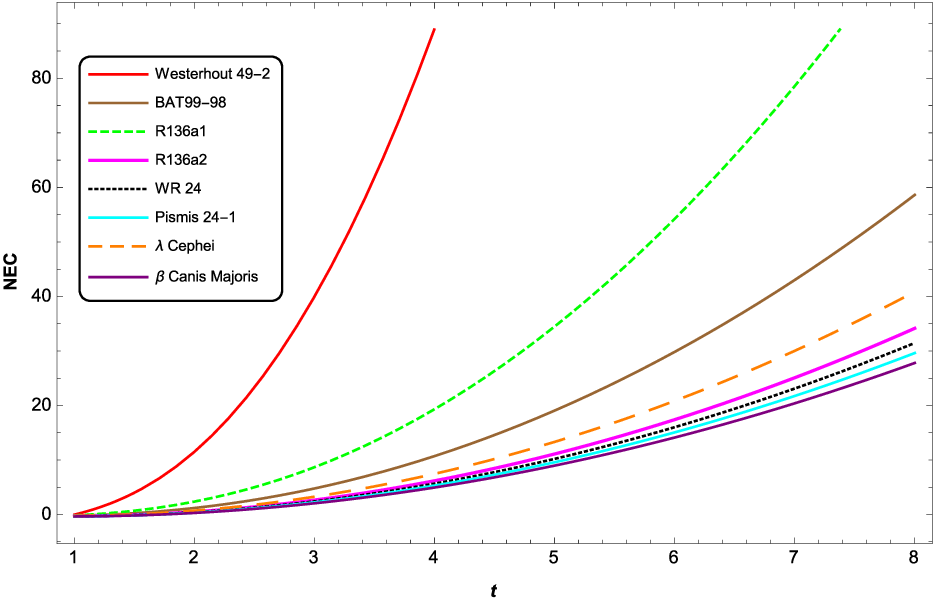}}
                \caption{}
			\label{figa}
		\end{subfigure}
  \hspace{1.1cm}
		\begin{subfigure}[a]{0.4\textwidth}
			{\includegraphics[width=8cm]{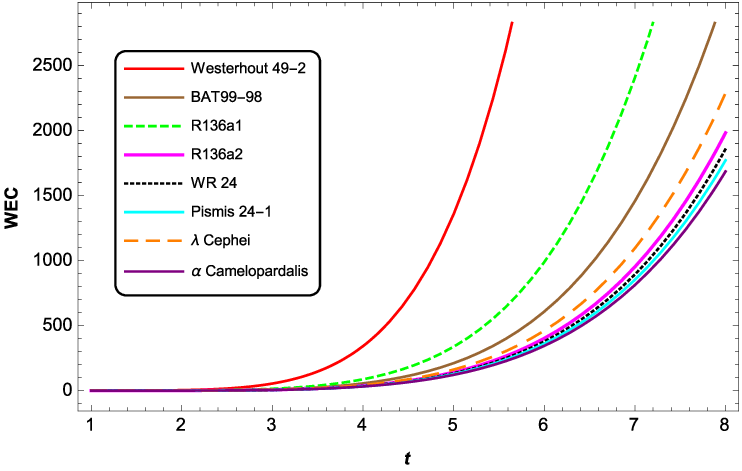}}
                 \caption{}
			\label{figb}
		\end{subfigure}
  
		\begin{subfigure}[c]{0.4\textwidth}
			{\includegraphics[width=8cm]{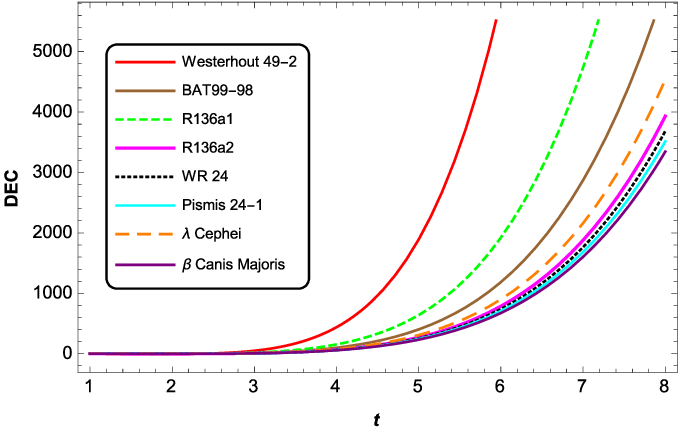}}
                 \caption{}
			\label{figc}
		\end{subfigure}
  \hspace{1.1cm}
		\begin{subfigure}[c]{0.4\textwidth}
			{\includegraphics[width=8cm]{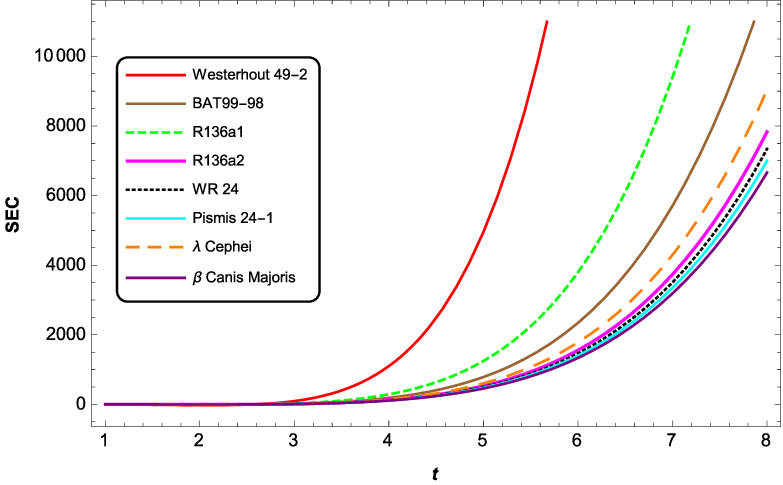}}
                  \caption{}
			\label{figd}
		\end{subfigure}
	\caption{ Physical evolution of energy conditions with time $t$ for model-1}
	\label{fig8}
\end{figure}

 \begin{figure}[hbt!]
		\begin{subfigure}[a]{0.4\textwidth}
			{\includegraphics[width=8cm]{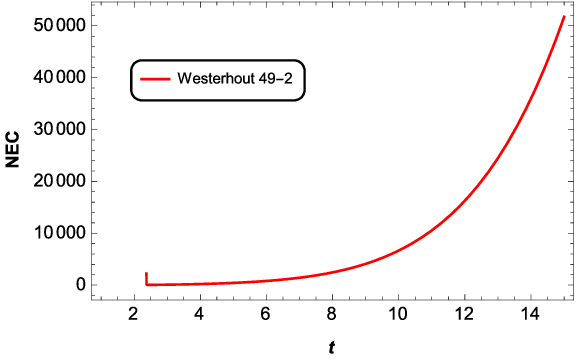}}
                \caption{}
			\label{fige}
		\end{subfigure}
  \hspace{1.1cm}
		\begin{subfigure}[a]{0.4\textwidth}
			{\includegraphics[width=8cm]{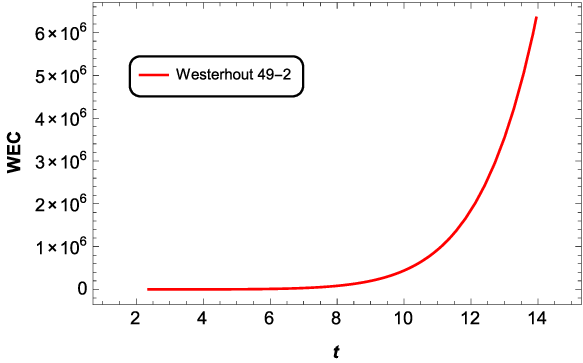}}
                 \caption{}
			\label{figf}
		\end{subfigure}
  
		\begin{subfigure}[c]{0.4\textwidth}
			{\includegraphics[width=8cm]{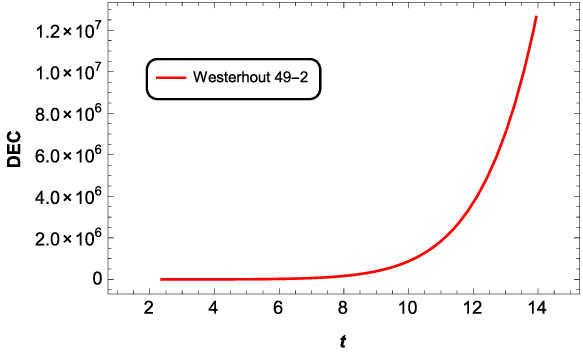}}
                 \caption{}
			\label{figg}
		\end{subfigure}
  \hspace{1.1cm}
		\begin{subfigure}[c]{0.4\textwidth}
			{\includegraphics[width=8cm]{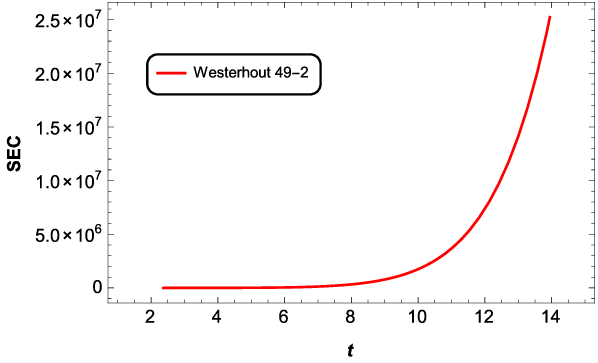}}
                  \caption{}
			\label{figh}
		\end{subfigure}
	\caption{ Physical evolution of energy conditions with time $t$ for model-2}
	\label{fig9}
\end{figure}
Figures \ref{fig8} and \ref{fig9} depict the physical behavior of various energy conditions given in Eqs.(\ref{ec1})–(\ref{ec4}) for our collapsing stellar models. It can be eminently seen that the trend of the energy conditions in both figures with their respective $f(R)$ models is physically valid and acceptable, which actually shows the potential viability of our model. Moreover, all the energy conditions reveal positive evolution and a regular nature at all points inside the stellar configuration.


\subsection{Stability Analysis of $f(R)$ Models}\label{sec11}
This section explores the stability of our models during stellar evolution, focusing on $f(R)$ gravity models that describe the interior nature of collapsing stars. Thus, the stability of collapsing stellar systems plays a valuable role in theoretical astrophysics. The cracking method is a potential method for analyzing the stability of the stellar model. Herrera\cite{Herrera1992} studied the cracking method for fluid distribution in self-gravitating stellar compact objects, determining the stability of collapsing fluid distribution. The condition suggests that for any stellar model to be considered physically acceptable, the speed of sound must satisfy causality conditions $0 < v_s^2 < 1$ i.e., the speed of sound propagation inside the star must be less than the speed of light $c$ (taken $c = 1$), where $v_s$ is the speed of sound propagation inside the star and is given by
\begin{equation}
		v^2_{s}= \lvert\frac{dp}{d\rho}\lvert
		\label{eq37}
	\end{equation}
For \textbf{Model-1:} Using Eqs.(\ref{eq27})-(\ref{eq28}) in (\ref{eq37}), one obtain
\begin{multline}
		v^2_{s}= \frac{1}{8 \gamma  t^3 \left[-4500 \alpha + t^2 \Bigl(-375 + 64   t \bigl(20 \alpha  t \left(15 + 2   t (15+ 8 \gamma  t)\gamma\right)  -t (35 + 16 \gamma  t)\gamma-25 \bigr)\gamma\Bigr) \right]} \Bigl[5625 t  \\ -3000 t^2\Bigl(72 \alpha \left(t^3+204 \alpha  t-9\right)\Bigr) \gamma + 3200 t^3 \Bigl(6 \alpha  (8 t^3-75) t-4 t^3+15\Bigr) \gamma ^2     \\ + 2560  t^4 \Bigl(60 \alpha  \left(2 t^3-9\right) t-7 t^3+15\Bigr)\gamma ^3  +4096  t^5 \left(40 \alpha  \left(t^3-3\right) t-2 t^3+3\right) \gamma ^4\Bigr]
		\label{eqs37a}
	\end{multline}

For \textbf{Model-2:} Using Eqs.(\ref{eq29})-(\ref{eq30}) in (\ref{eq37}), one obtain
 \begin{multline}
		v^2_{s}= \frac{1}{\mathcal{D}}  \Biggl[-54 \beta ^3 \Bigl(8 \gamma \left(8 \gamma  \mathcal{G} t+60 t^3-135\right)t-675\Bigr)^3- 16 \gamma  \Bigl(16 \gamma  t \left(2 t^3-9\right)-135\Bigr)^3  t^4 \times \\ \Bigl(-225 + 8 \gamma   \left(8 \gamma   (2 t^3-3) t+15 (t^3-3)\right)t\Bigr)+ 3 \beta ^2 \Bigl(-675 +8 \gamma   (-135 + 8 \gamma \mathcal{G} t + 60 t^3)t\Bigr) \times \\ \Biggl(12301875+16 \gamma  t \Bigl(-5400 \gamma  t (2 t^3-135) (4 t^3+9)+ 455625 (2 t^3+9)+\\ 512 \gamma ^3  \left(64 t^9+72 t^6+2268 t^3+2187\right) t^3 +1920 \gamma ^2  \left(16 t^9+36 t^6+1701 t^3+2187\right) t^2 \Bigr)\Biggr)+ \\ \beta  \Biggl(8303765625+ 8 \gamma t  \Bigl(-184528125 (13 t^3-45)-10935000 \gamma  t (46 t^6+1215 t^3-2511)- \\ 230400 \gamma ^3 t^3 \left(304 t^{12}+792 t^9+27621 t^6+139239 t^3-157464\right)+\\ 245760 \gamma ^4 t^4 \Bigl(2 \left(2 t^3 \left(8 t^9-162 t^6-324 t^3-5103\right)-32805\right) t^3+59049\Bigr)- 5832000 \gamma ^2 t^2 \\ \left(8 t^9+566 t^6+5139 t^3-7614\right)+ 131072 \gamma ^5 t^5  \times \\ \left(64 t^{15}-648 t^{12}-864 t^9-10206 t^6-26244 t^3+19683\right)\Bigr)\Biggr)\Biggr]
		\label{eqs37b}
	\end{multline}
 where $\mathcal{G}= 8t^3 -9$ and $D= 8 \gamma  t^4 \Biggl(18 \beta ^2 \left(16 \gamma t \left(2 t^3-9\right)-135\right) \left(675-8 \gamma t \left(8 \gamma  \left(8 t^3-9\right) t+60 t^3-135\right)\right)^2- \\ 16 \gamma  t^4 (16 \gamma  t+15) \left(16 \gamma t \left(2 t^3-9\right)-135\right)^3+\beta  \Bigl(553584375+16 \gamma t \Bigl(12301875 \left(t^3+18\right)+243000 \gamma t \left(52 t^6+207 t^3+2025\right)-388800 \gamma ^2 t^2  \left(t^3 \left(16 t^3 \left(t^3-7\right)-261\right)-1296\right) + 7680 \gamma ^3 t^3 \left(8 \left(8 t^9-234 t^6+729 t^3+1458\right) t^3+32805\right) + \\ 8192 \gamma ^4 t^4 \left(2 t^3-9 \right) \left(32 t^9-324 t^6-486 t^3-729\right)\Bigr)\Bigr)\Biggr)$

 \par
\begin{figure}[hbt!]
\subfloat[Model 1]{
	\begin{minipage}[a]{0.4\textwidth}
		{\includegraphics[width=8.7cm]{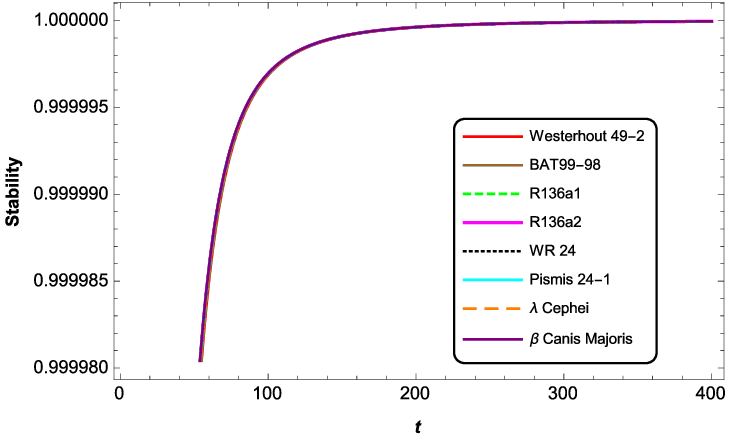}}
		\label{fige}
	\end{minipage}}
 \hspace{1.8cm}
	\subfloat[Model 2]{
	\begin{minipage}[a]{0.4\textwidth}
 \centering
		{\includegraphics[width=8.2cm]{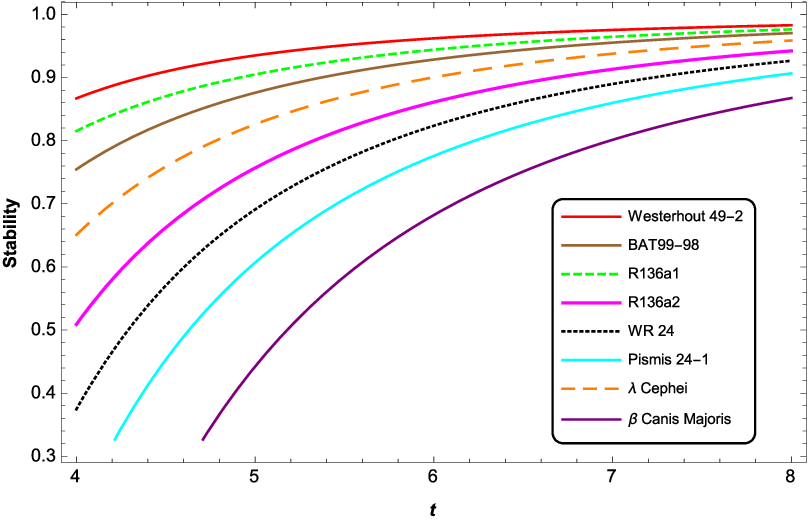}}
		\label{figf}
	\end{minipage}}
 \caption{Stability analysis for both models}
 \label{stability}
 \end{figure}

 As a result,  we have determined the sound's speed($v_s$) for both the models and it can be observed in figure-\ref{stability}  that $v_s^2 <1 $ throughout the evolution of the stellar system\footnote{For the model-1, all the plots of $v_s^2$ for eight massive stars are overlapping.}. Hence, one can conclude that both $f(R)$ models satisfy the stability criteria.  

	\section{Eternal Collapse Phenomenon: Singularity analysis}\label{sec6}
	An Eternally Collapsing Object (ECO) is a hot, compact star that collapses due to its intense gravitational pull. Due to their compactness, it resemble "mathematical black holes," but observational evidence suggests that these alleged astrophysical objects may called ECOs. According to GR, space-time is like a fabric that stretches, curves, and forms a pit with the increase of gravity; an object undergoing collapse has to slide down this membrane into the pit. If gravity's strength increases continuously during collapse ($t\rightarrow \infty$), the gravitational pit may become endless and bottomless. Since the stellar object would become hotter and hotter under gravitational compression, the pressure becomes so strong that it fails to counterbalance the inward pull of gravity\cite{Robertson2010} \cite{Leiter2003}.  If so, continued gravitational collapse would indeed become eternal without the formation of any real space-time singularity.  Thus, the stellar objects undergoing continued general relativistic collapse would end up as “Eternally Collapsing/Contracting Objects” rather than as true Black Holes or “Naked Singularities”.
	\par
During gravitational collapse in stellar objects, the formation of apparent horizons (trapped surfaces) in space-time characterizes the various outcomes as either a black hole (BH) or a naked singularity (NS) \cite{book1}. When a star begins to collapse owing to its own gravity, no portions of space-time are initially trapped; nevertheless, after sufficiently high density is reached, trapped surfaces emerge and an apparent horizon region develops. It is speculated that self-gravitating stellar objects often wind up with the development of a space-time singularity as a consequence of gravitational collapse, which is defined by divergence curvature and energy density.
\par 

For the space-time metric (\ref{eq1}), the apparent horizon is characterized by \cite{EPJC}\cite{Bhattacharjee2018}
\begin{equation}
	\mathcal{R}_{,i} \mathcal{R}_{,j} g^{ij} = \dot{\mathcal{R}}^2(t,r) - 1 = 0
	\label{eq39}
\end{equation} 
Since the present study is concerned with the formation of space-time singularity due to the gravitational collapse of the star, we assume that at the initial of the collapse $(t_{0},r_{0})$ the star is not trapped that is, 
\begin{equation}
	\mathcal{R}_{,i} \mathcal{R}_{,j} g^{ij}|_{(t_{0},r_{0})} = r_{0}^2 \dot{a}^2(t_{0}) - 1 < 0
	\label{eq399}
\end{equation}
Let us assume that the collapsing star forms an apparent horizon surface at $(t_{AH},r_{AH})$, then it follows from  (\ref{eq39}) that 
\begin{equation}
	\dot{\mathcal{R}}^2(t_{AH},r_{AH})-1 = 0
	\label{eq400}
\end{equation}
Recently authors\cite{EPJC} have obtained that $t_{AH}$, time formation of the apparent horizon is finite, which is given by
\begin{equation}
	\begin{split}
		\frac{8 \sqrt[3]{\frac{2}{3}} \gamma ^{\frac{2}{3}} M^{\frac{2}{3}} r_{AH}^2 t_{AH}^6 }{3 r_0^2 t_0^4} (8 \gamma  t_0 +5)^{\frac{4}{3}-\frac{125}{1152 \gamma^3}} (8 \gamma  t_{AH}+5)^{\frac{125}{1152 \gamma^3}-2} &\\ e^{ \frac{64 \gamma ^2 t_0^3-60 \gamma  t_0^2+75 t_0+t_{AH} \left(-64 \gamma^2 t_{AH}^2+60 \gamma  t_{AH}-75\right)}{432 \gamma ^2}} = 1		
		\label{eq42}
	\end{split}
\end{equation}

\textbf{Singularity analysis}
\par
In NS, the trapped surface forms at the center of the cloud at the time of the formation of singularity, and the apparent horizon then moves outwards to meet the event horizon at the boundary at a time later than singularity formation \cite{Bhattacharjee2018}. In the BH scenario, the trapped surface emerges earlier than the singularity formation. The outside event horizon covers the final stages of collapse when a singularity forms, while the apparent horizon inside matter evolves from the outer shell to the singularity\cite{Bhattacharjee2018}- \cite{Ellis2003}.
\par
 Assuming that the star begins to collapse at $t=t_{0}$ where condition (\ref{eq399}) holds i.e., the star is not initially trapped. From Eqs.(\ref{eq25}), (\ref{eq27}), and (\ref{eq29}) one can see that  $\mathcal{K}\rightarrow \infty $, $\rho \rightarrow \infty $ as $t\rightarrow \infty $, i.e., the Kretschmann curvature and energy density diverge over infinite comoving time, leading the star to collapse for an infinite duration to reach the space-time singularity.
 \par
 
 Since $t_{AH} < \infty$, therefore the singularity is not naked as an apparent horizon is already formed at $t_{AH}$ before it is formed. Further one can see from eq. (\ref{eq22}) that mass($m$) vanishes as $t\rightarrow \infty $ \cite{EPJC}. Since, both the BH mass and the comoving time for its formation must be finite, i.e., the BH singularity must be formed during gravitational collapse with finite mass in a finite time rather than $m\rightarrow 0$ and $t\rightarrow \infty $ and therefore, the true BH is not formed here. It should be noticed here that in GR, $ m \rightarrow 0$ event doesn't imply the absence of matter, as gravitational mass comprises all energy sources, including negative self-gravitational energy, and such phenomenon may indicate extreme self-gravitation, offsetting other energy sources like protons, neutrons, and internal energies like heat and pressure. Also, from eq.(\ref{eq26}) one can see that acceleration continuously increases showing an accelerating phase of gravitational collapse (see figure-\ref{fig4}). Thus with the singularity analysis of our models, one may conclude that stellar systems tend to collapse for infinite comoving time in order to attain the singular state $(\mathcal{K}\rightarrow \infty, \rho \rightarrow \infty)$ and therefore may be called Eternal Collapsing phenomenon\cite{Mitra2010}\cite{EPJC}.
 
\section{Discussion and Concluding Remarks}\label{sec12}
The present study aims to discuss some new aspects of the spherically symmetric, homogeneous gravitational collapsing phase of the stellar system (Eternal collapse) in $f(R)$ gravity theory and provides an appropriate model that incorporates known astrophysical stellar data.  The exact solutions and singularity analysis studies are crucial in $f(R)$ theory, and very few models provide physically interesting results in an astrophysical scenario. The interior space-time of the system is considered to be the FLRW metric with a perfect fluid distribution. Moreover, we have smoothly connected the interior line element with the exterior Schwarzschild metric at the 3-dim hypersurface $\Sigma$ to extract the dynamical evolution equation. By using the generalized Darmois-Israel junction conditions for $f(R)$ gravity theory, we have obtained the required boundary conditions for the smooth matching of the interior and exterior spacetimes. Since $\Theta^{-1}= H^{-1}$ is the only natural time scale in the model, we chose a parametrization of the expansion scalar of the form (\ref{eq15}), where $H$ is the cosmological Hubble parameter. This parametrization allowed us to solve the governing field equations, which subsequently determined the gravitational and dynamical evolution of the collapse process.  In order to assess the astrophysical relevance of our model in terms of graphical depiction, we have taken into consideration the massive stars, namely, $Westerhout 49-2, BAT99-98, R136a1, R136a2, WR 24, Pismis 24-1$, $\lambda -Cephei$,  and $\beta -Canis Majoris$ and throughout the work, all graphs are drawn for these stellar candidates. We expect that the physics of the stellar model would be sensitive to the model parameters  $\alpha,\beta$, and $\gamma$, hence in our study we have calculated them numerically for the known data points (their masses and radii) of the aforementioned stars. We have manifested that the model is in good agreement with and compatible with various physical conditions of stability and is well consistent under two $f(R)$ models, namely; $f(R) = R+\alpha R^2$ and $f(R) = R^{1+\beta}$. With the critical singularity analysis of our stellar models, we observe that no real singularity (BH or NS) forms throughout the collapsing configurations, though it reveals an eternal collapsing phenomenon (ECO) of the star. Apart from these feasible results, we have also checked the physical credibility of our models via various criteria, namely energy conditions and stability analysis.
\par
Based on the aforementioned discussion, the proposed $f(R)$ models of collapsing stellar systems have highlighted the following various salient features (The unit of time $t$  throughout the graphical representations is in Myr (Million years)):

	\par
$\bullet$ \textbf{Physical quantities:} The study explores the evolution of energy density and pressure through graphical representation in two feasible $f(R)$ models. It can be evidently seen from Eqs. (\ref{eq27})-(\ref{eq30})  that both are finite and regular at all interior points of the stellar configuration, and both are increasing in nature \ref{fig4a}- \ref{fig7}, where the negative sign of $p$ indicates the direction of pressure towards the center ($r = 0$) during the collapsing configuration.

\par
$\bullet$ \textbf{Kinematical quantities:} Expansion scalar $\Theta$ implies the collapsing nature of the star which happens due to the nuclear fusion at the core of the star. The data shows that it increases over time and tends to infinity at the center $r = 0$, as depicted in Figure-\ref{fig1}, where the negative sign of $\Theta$ represents the direction of motion of the concentric fluid shells toward the stellar center.  The scale factor $a$ and radius $\mathcal{R} (= ra)$ are regular and finite inside the system and decrease as the collapse proceeds. This trend is depicted in   figure-\ref{fig2}. Eq.(\ref{eq19}) reveals that the collapse reaches its central singularity ($a \rightarrow 0$) as $t \rightarrow \infty$. The increase in collapse acceleration over time indicates the accelerating phase of collapse in the stellar system as shown in figure-\ref{fig4}.

\par
$\bullet$ \textbf{Energy conditions and stability:} The physical viability of the $f(R)$ stellar models was investigated under different energy conditions(NEC, WEC, SEC , DEC) in Figure-\ref{fig8} and \ref{fig9}. All energy conditions are consistently met and validated at all points throughout the collapsing configuration. We have investigated the stability analysis of our models, which implies that our solutions have a well-behaved nature within the star as shown in figure-\ref{stability}. Thus, our proposed $f(R)$ models suggest a viable collapse scenario of the stellar model, with physically realizable astrophysical implications.
\par
$\bullet$ \textbf{Eternal Collapse (no real singularity):} The singularity analysis was conducted by comparing the time of singularity formation and the time of apparent horizon formation ($t_{AH}$). The Kretschmann curvature and energy density exhibit divergent behavior as $t \rightarrow \infty$, with a singularity attributed to gravitational collapse in infinite co-moving time. Therefore, the gravitational collapse is likely to end in quasi-stable, ultra-compact, or continuing configurations. Thus, the continued gravitational collapse should lead to an eternal state, without real singularity, as objects become eternally collapsing or contracting, rather than black holes or naked singularities.
\par

This work is ground-breaking as it presents a consistent general relativistic $f(R)$ model for describing the continued collapsing scenarios of stellar objects with their known masses and radii. Hence, it demonstrates that our suggested model is highly significant for realistic stellar systems. It is also noticeable that our models accurately describe physically realizable stellar structures without introducing exotic matter distributions like dark energy and dark matter. Future studies should consider generalizing the conditions for a perfect fluid to include anisotropic stresses, shear, density inhomogeneity, and dissipation in the stellar interior. It would be interesting to determine the role of relaxational effects brought in through causal heat transport, especially during the late stages of gravitational collapse. 

\par
	\textbf{Acknowledgment:} The authors AJ, RK, and SKS are acknowledged to the Council of Science and Technology, UP, India vide letter no. CST/D-2289.	 Author RK is also thankful to IUCAA for all their support, where a part of the work done during the visit.
	

\end{document}